\documentclass[aps,preprint,amsmath,amssymb,showpacs]{revtex4}

\usepackage{graphicx}
\begin{document}

\title{Unparticle Effects on Top Quark Spin Correlations in
$e^{+}e^{-}$ Collision}

\author{Banu \c{S}ahin}
\email[]{dilec@science.ankara.edu.tr} \affiliation{Department of
Physics, Faculty of Sciences, Ankara University, 06100 Tandogan,
Ankara, Turkey}

\begin{abstract}

We investigate the effects of scalar and vector unparticles on top
quark spin correlations via the process $e^{+}e^{-}\rightarrow t
\overline{t}$. In addition to the Standard Model diagrams, there is
a new contribution to top-antitop quark production process mediated
by unparticle in the s-channel. It is shown that scalar and vector
unparticle contribution leads to a considerable deviation of the top
spin correlations from the Standard Model one.

\end{abstract}


\maketitle

\section{Introduction}

There has been an increasing interest in unparticle scenario which
was introduced by Georgi \cite{Georgi, Georgi2}. In this scenario,
new physics contains both Standard Model (SM) fields and a scale
invariant sector described by Banks-Zaks (BZ) fields \cite{BZ}.
These two sector interact via the exchange of particles with a mass
scale $M_{U}$. Below this large mass scale interactions between SM
fields and BZ fields are described by non-renormalizable couplings
suppressed by powers of $M_{U}$ \cite{Georgi, Cheung}:

\begin{eqnarray}
\frac{1}{M_{U}^{d_{SM}+d_{BZ}-4}}O_{SM}O_{BZ}
\end{eqnarray}

Renormalization effects in the scale invariant BZ sector then
produce dimensional transmutation at an energy scale $\Lambda_{U}$
\cite{Weinberg}. In the effective theory below the scale
$\Lambda_{U}$, the BZ operators are embedded as unparticle
operators. The operator (1) is now match onto the following form,

\begin{eqnarray}
C_{O_{U}}\frac{\Lambda_{U}^{d_{BZ}-d_{U}}}{M_{U}^{d_{SM}+d_{BZ}-4}}O_{SM}O_{U}
\end{eqnarray}
here, $d_{U}$ is the scale dimension of the unparticle operator
$O_{U}$ and the constant $C_{O_{U}}$ is a coefficient function.

Phenomenological implications of unparticles have been discussed in
the literature \cite{Luo}. In some of these researches mentioned
above several unparticle production processes have been studied. A
possible evidence for this scale invariant sector might be a missing
energy signature. It can be tested experimentally by examining
missing energy distributions. Another evidence for unparticles can
be explored by studying its virtual effects. Imposing scale
invariance, unparticle propagators for spin-0 and spin-1 unparticles
are given by \cite{Georgi2,Cheung2}:

\begin{eqnarray}
\Delta(P^{2})&&=i\frac{A_{d_{U}}}{2sin(d_{U}\pi)}(-P^{2})^{d_{U}-2}
\nonumber \\
\Delta(P^{2})^{\mu\nu}&&=i\frac{A_{d_{U}}}{2sin(d_{U}\pi)}(-P^{2})^{d_{U}-2}
\left(-g^{\mu\nu}+\frac{P^{\mu}P^{\nu}}{P^{2}} \right)
\end{eqnarray}

respectively, where

\begin{eqnarray}
A_{d_{U}}=\frac{16\pi^{\frac{5}{2}}}{(2\pi)^{2d_{U}}}\frac{\Gamma(d_{U}+\frac{1}{2})}
{\Gamma(d_{U}-1)\Gamma(2d_{U})}
\end{eqnarray}

In Eq.(3), $(-P^{2})^{d_{U}-2}=|P^{2}|^{d_{u}-2}$ in t or u-channel
diagrams where $P^{2}$ is negative,
$(-P^{2})^{d_{U}-2}=|P^{2}|^{d_{u}-2}e^{-id_{u}\pi}$ in s-channel
diagrams where $P^{2}$ is positive.

Interaction vertices for the scalar and vector unparticles with the
SM fermions are given respectively, by

\begin{eqnarray}
i\frac{\lambda_{0}}{\Lambda_{U}^{d_{U}-1}}-\frac{\lambda_{0}}{\Lambda_{U}^{d_{U}-1}}\gamma^{5}+
\frac{\lambda_{0}}{\Lambda_{U}^{d_{U}}}\gamma^{\mu}p_{\mu}
\end{eqnarray}
\begin{eqnarray}
i\frac{\lambda_{1}}{\Lambda_{U}^{d_{U}-1}}\gamma^{\mu}+i\frac{\lambda_{1}}{\Lambda_{U}^{d_{U}-1}}\gamma^{\mu}
\gamma^{5}
\end{eqnarray}

The top quark is the heaviest fermion in the Standard Model (SM) and
its mass is at the electroweak symmetry-breaking scale. Because of
its large mass top quark couplings are expected to be more sensitive
to new physics than other particles\cite{Peccei}. Therefore
deviations from the SM expectations in top quark production
processes would be a signal for the new physics.

In this work we analyzed vector and scalar unparticle effects on top
quark spin correlations in pair production process
$e^{+}e^{-}\rightarrow t\overline{t}$. Since the top quark is very
heavy its weak decay time is much shorter than the typical time for
the strong interaction to affect its spin\cite{Bigi}. Therefore top
polarization information is not distributed by hadronization effects
and transferred to the decay products. The angular distribution of
the top quark decay involves correlations between top decay products
and top quark spin:

\begin{eqnarray}
\frac{1}{\Gamma_{T}}\frac{d\Gamma}{dcos\theta}=\frac{1}{2}(1+A_{\uparrow\downarrow}\alpha
cos\theta)
\end{eqnarray}

Here the dominant decay chain of the top quark in the standard model
$t \to W^{+}b(W^{+} \to l^{+}\nu,\bar{d}u)$ is considered.
$A_{\uparrow\downarrow}$ is the spin asymmetry and $\theta$ is
defined as the angle between top quark decay products and the top
quark spin quantization axis in the rest frame of the top quark.
$\alpha$ is the correlation coefficient and $\alpha=1$ for $l$ or
$\bar{d}$ which leads to the strongest correlation. Therefore top
quark polarization can be determined by means of the angular
distribution of its decay products.

We take into account top quark spin and antitop quark spin
polarizations along the direction of various spin bases. These spin
bases are the helicity basis and the incoming beam directions. In
the SM there is a spin asymmetry between the produced top-antitop
pairs. More specifically, the number of produced top-antitop quark
pairs with both spin up or spin down is different from the number of
pairs with the opposite spin combinations. Therefore, if the top
quark is coupled to a new physics beyond the SM, the top-antitop
quark spin correlations could be altered\cite{smolek}.

The top-antitop correlations can be used to search new physics
beyond the SM. The unparticle couplings are examples of such new
physics. We consider the scalar and vector unparticle interaction
terms which are given in Eq.5,6 in addition to the SM contributions.
In our calculations we assume that $\lambda_{0}=\lambda_{1}=1$.

The research and developments on linear $e^{+}e^{-}$ colliders have
been progressing and physics potential of these future machines is
under study. Linear $e^{+}e^{-}$ collider can provide a very useful
laboratory to study physics of the top quark. Furthermore this
linear colliders have a clean environment and the experimental
clearness is an additional advantage of $e^{+}e^{-}$ collisions with
respect to hadron collisions.

\section{Top Quark Pair Production and Spin Correlations}

Since top quark possesses a large mass its helicity is frame
dependent and changes under a boost from one frame to another. The
helicity and chirality states do not coincide with each other and
there is no reason to believe that the helicity basis will give the
best description of the spin of top quarks. Therefore it is
reasonable to study other spin bases for top quark.

Top quark spinors are eigenstates of the operator
$\gamma_{5}(\gamma_{\mu}s_{t}^{\mu})$:

\begin{eqnarray}
[\gamma_{5}(\gamma_{\mu}s_{t}^{\mu})]u(p_{t},\pm s)=\pm u(p_{t},\pm
s)
\end{eqnarray}
where spin four vector of a top quark is defined by

\begin{eqnarray}
s_{t}^{\mu}=(\frac{\overrightarrow{p_{t}}.\overrightarrow{s'}}{m_{t}},\overrightarrow{s'}+
\frac{\overrightarrow{p_{t}}.\overrightarrow{s'}}{m_{t}(E_{t}+m_{t})}\overrightarrow{p_{t}})
\end{eqnarray}
here $(s_{t}^{\mu})_{RF}=(0,\overrightarrow{s'})$ in the top quark
rest frame.

In the presence of unparticles, $e^{+}e^{-}\rightarrow
t\overline{t}$ process occur via s-channel exchange of unparticles
and usual electroweak bosons, $\gamma$ and Z. During amplitude
calculations one can project the top quark and antitop quark spin to
a given spin direction. We consider incoming beam directions and
helicity basis. In the top quark rest frame (or in antitop quark
rest frame),its spin direction along any beam can be defined as,

\begin{eqnarray}
\overrightarrow{s'}=\lambda\frac{\overrightarrow{p^{*}}}{|\overrightarrow{p^{*}}|},\,\,\,\,\,
\lambda=\pm1.
\end{eqnarray}
where, $\overrightarrow{p^{*}}$ is the particle momentum (positron
or electron), observed in the rest frame of the top quark.

It is possible to analyze the top and antitop quark spin
correlations via angular correlations of two charged leptons
$l^{+}l^{-}$ produced by the top-antitop quark leptonic decay
channels. We consider the leptonic decay channels of the top pairs.
After integration over azimuthal angles the correlation is given by
\cite{smolek,stelzer},

\begin{eqnarray}
\frac{1}{\sigma}\frac{d^{2}\sigma}{dcos\theta_{l^{+}}dcos\theta_{l^{-}}}=
\frac{1-\textit{A}cos\theta_{l^{+}}cos\theta_{l^{-}}}{4}
\end{eqnarray}

Here $\sigma$ represents the cross section for the process of the
leptonic decay modes, $\theta_{l^{+}}(\theta_{l^{-}})$ represents
the angle between the top (antitop) spin axis and the direction of
motion of the anti-lepton (lepton) in the top (antitop) rest frame.
The \textit{A} coefficient stands for the spin asymmetry between top
and antitop pairs and defined as,

\begin{eqnarray}
\textit{A}=\frac{\sigma(t_{\uparrow}\overline{t}_{\uparrow})+
\sigma(t_{\downarrow}\overline{t}_{\downarrow})-\sigma(t_{\uparrow}\overline{t}_{\downarrow})-
\sigma(t_{\downarrow}\overline{t}_{\uparrow})}{\sigma(t_{\uparrow}\overline{t}_{\uparrow})+
\sigma(t_{\downarrow}\overline{t}_{\downarrow})+\sigma(t_{\uparrow}\overline{t}_{\downarrow})+
\sigma(t_{\downarrow}\overline{t}_{\uparrow})}
\end{eqnarray}

where $\sigma(t_{\alpha}\overline{t}_{\alpha'})$ is the cross
section of pair top quark production process.

In Table \ref{tab1}-\ref{tab2} spin asymmetries are given for
various spin bases. One can see from Table \ref{tab1}, SM spin
asymmetry is $\textit{A}_{SM}=-0.651$ for $\sqrt{s}$=0.5 TeV when
top spin and antitop quark spin are in the helicity basis. In
unparticle scheme, there are new contributions to pair production
process. We calculated spin dependent squared amplitudes including
scalar and vector unparticle mediated process. In the scalar
unparticle exchange case, spin asymmetry increases and it takes a
positive value, $\textit{A}_{S}=0.986$. When we consider vector
unparticle exchange, asymmetry takes a negative value,
$\textit{A}_{V}=-0.727$. It is seen from the tables that maximum
deviation from the SM one occurs in the scalar unparticle case. On
the other hand, another useful basis is the incoming beam
directions, in which the top quark spin axis is along the positron
beam direction in the top rest frame, and antitop quark spin axis is
along the electron direction in the antitop rest frame. In this
basis, SM asymmetry is $\textit{A}_{SM}=-0.939$ for $\sqrt{s}$=0.5
TeV. In scalar unparticle exchange, $\textit{A}$ increases and takes
a positive value, $\textit{A}_{S}=0.657$. In vector unparticle
exchange spin asymmetry is, $\textit{A}_{V}=-1$. It is shown that
maximum deviation from the SM observed in the scalar unparticle
exchange for the helicity basis. At the tables we considered that
$d_{U}=1.1$ and $\Lambda_{U}=1$ TeV.

\begin{figure}
\includegraphics{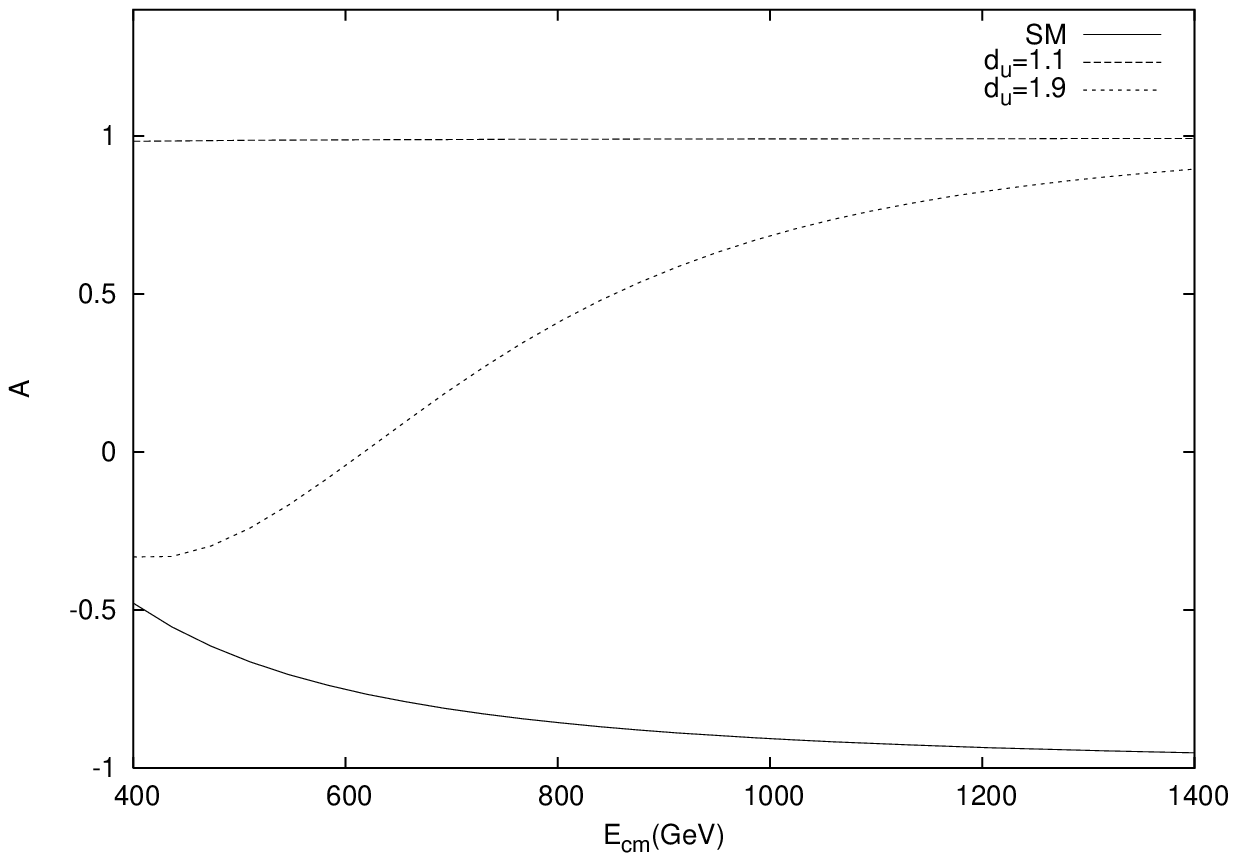}
\caption{Spin asymmetry as a function of center of mass energy in
the helicity basis for the scalar unparticle. The scale dimension is
$d_{U}=1.1, 1.9$. $\Lambda_{U}=1 TeV$ \label{fig1}}
\end{figure}

\begin{figure}
\includegraphics{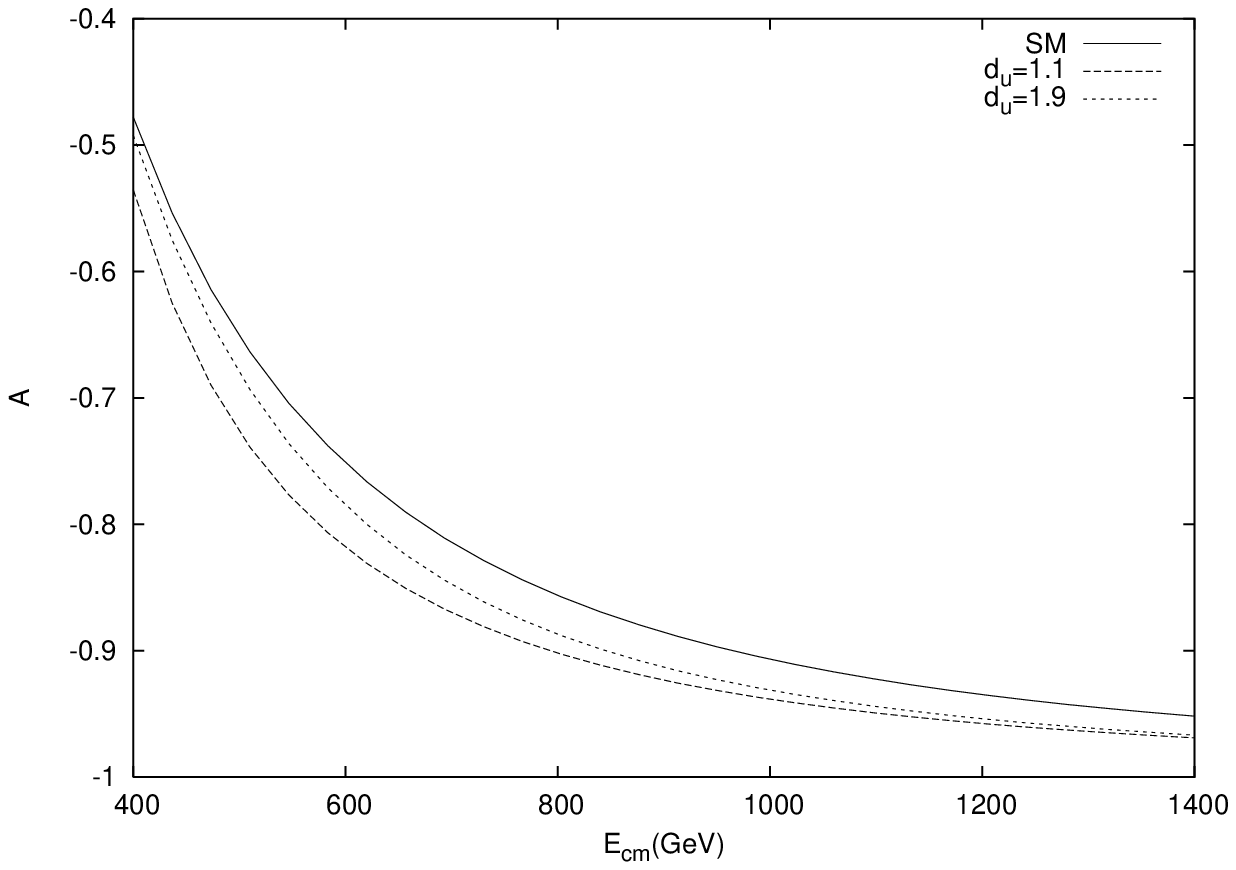}
\caption{The same as Fig1 but for the vector unparticle.
\label{fig2}}
\end{figure}

\begin{figure}
\includegraphics{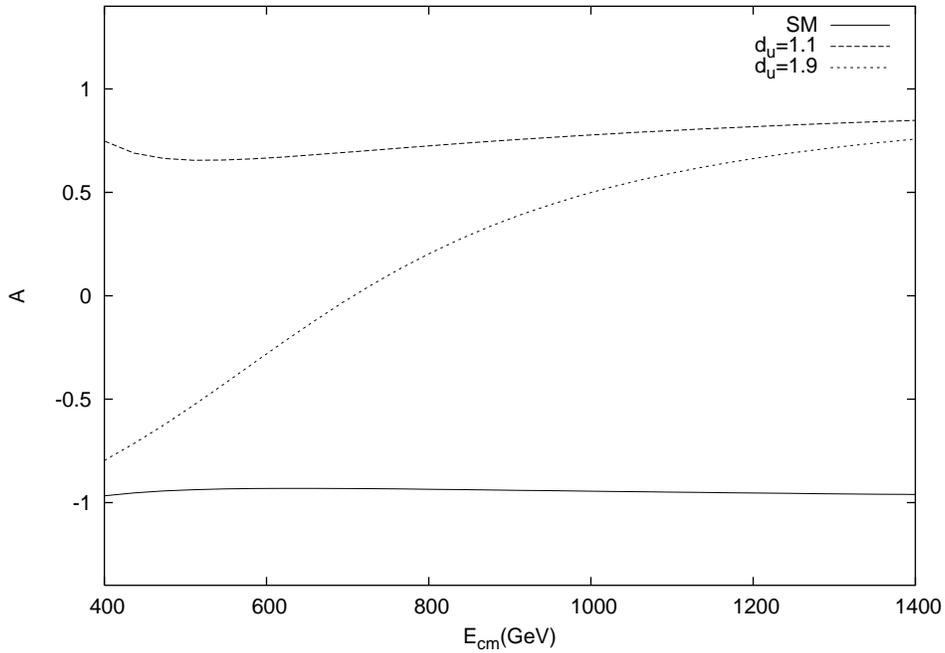}
\caption{Spin asymmetry as a function of center of mass energy for
the scalar unparticle. Top quark spin axis is along the positron
beam direction and antitop quark spin axis is along the electron
beam direction. The scale dimension $d_{U}=1.1, 1.9$. $\Lambda_{U}=1
TeV$. \label{fig3}}
\end{figure}

\begin{figure}
\includegraphics{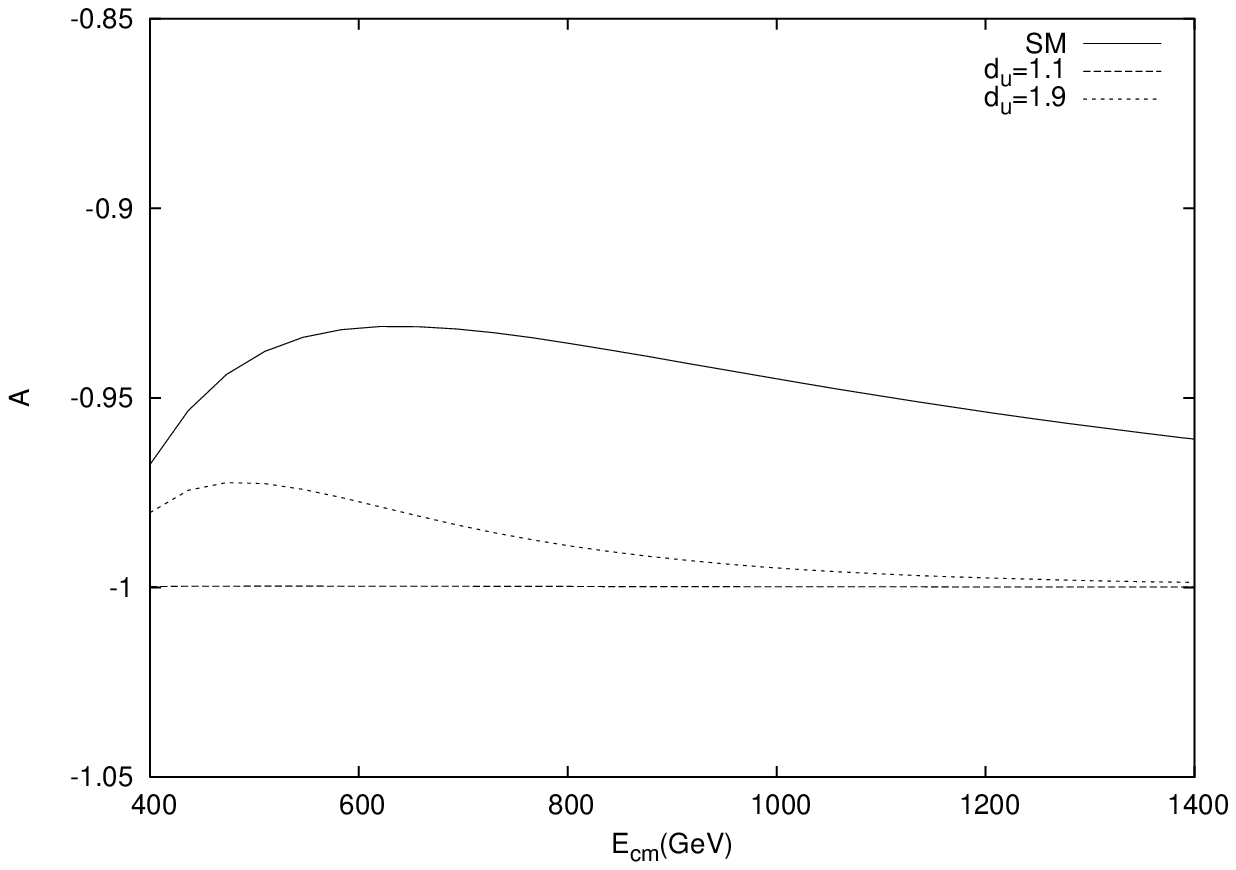}
\caption{The same as Fig3 but for the vector unparticle.
\label{fig4}}
\end{figure}

The influence of the center of the mass energy on the spin asymmetry
is shown in Fig.\ref{fig1}-\ref{fig4}. In Fig.\ref{fig1} we consider
the scalar unparticle and helicity basis for top-antitop quark spin.
We see from this figure that a sizeable deviation from the SM one
occurs at $d_{U}=1.1$. In Fig.\ref{fig2} we take into account vector
unparticle effects. One can see that unparticle contributions leads
to a little deviation from the SM at each center of mass energy.
Fig.\ref{fig3} shows similar behavior with Fig.\ref{fig1} but for
the top spin axis is along the incoming positron beam direction and
antitop spin axis is along the incoming electron beam direction. In
Fig.\ref{fig4} vector unparticle effects are shown for the same
basis with Fig.\ref{fig3}. In this figure maximum deviation can be
seen at $d_{U}=1.1$.

\begin{figure}
\includegraphics{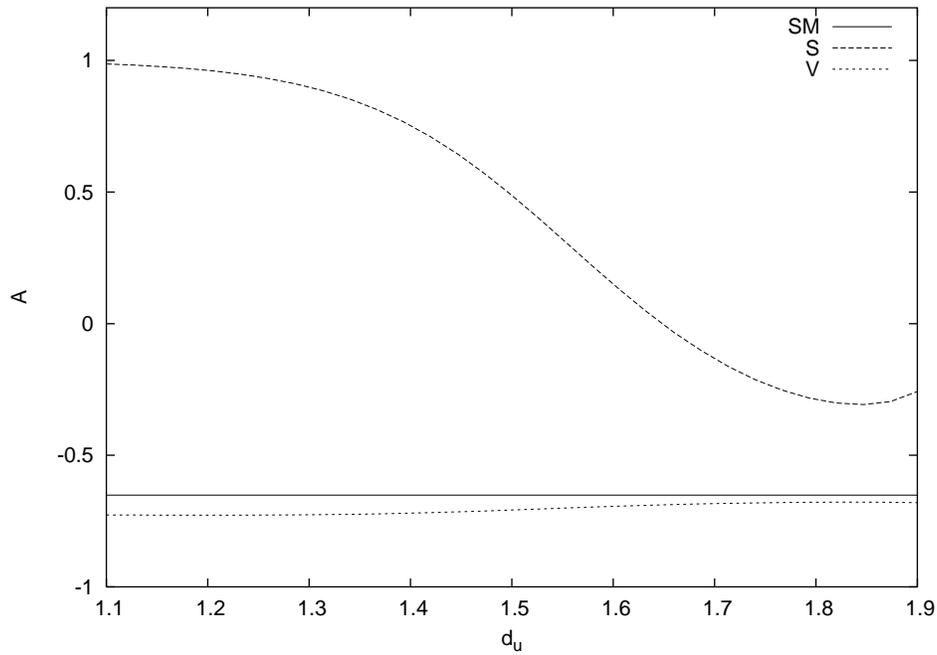}
\caption{Spin asymmetry as a function of scale dimension $d_{U}$ in
the helicity basis for the scalar and vector unparticle.
$\sqrt{s}=0.5$ TeV, $\Lambda_{U}=1 TeV$ \label{fig5}}
\end{figure}

\begin{figure}
\includegraphics{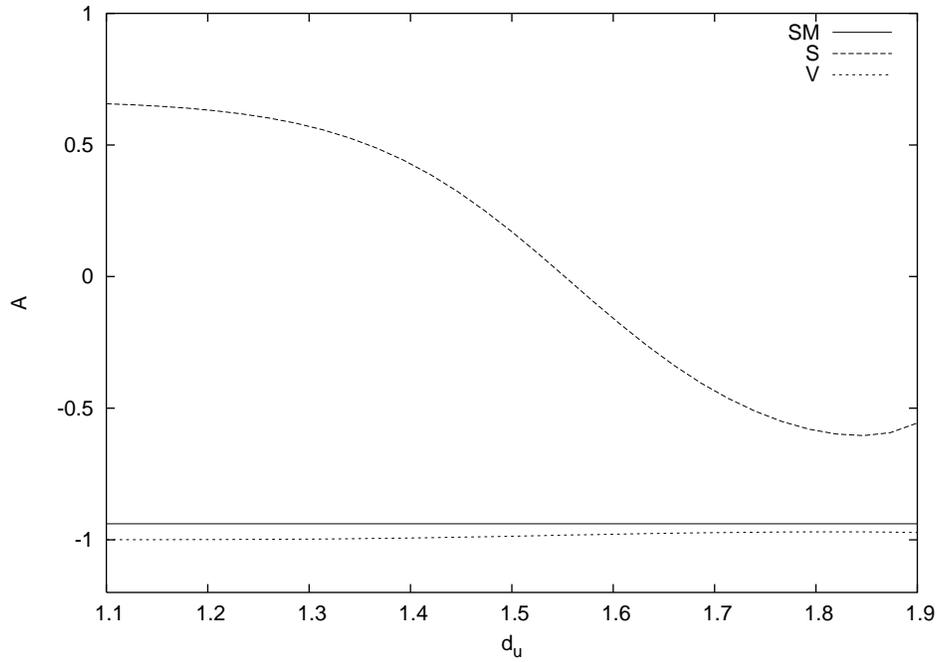}
\caption{Spin asymmetry as a function of scale dimension $d_{U}$ for
the scalar and vector unparticle. Top quark spin axis is along the
positron beam direction and antitop quark spin axis is along the
electron beam direction. $\sqrt{s}=0.5$ TeV, $\Lambda_{U}=1 TeV$
\label{fig6}}
\end{figure}

In Fig.\ref{fig5} the spin asymmetries as a function of the scale
dimension $d_{U}$ are given. We consider helicity direction in this
figure and it can be seen that scalar unparticle contribution leads
to a considerable deviation from the SM. On the other hand, vector
unparticle contribution leads to a little deviation. At growing
values of $d_{U}$ SM, scalar and vector unparticle  asymmetries
close up. Fig.\ref{fig6} has similar behavior with Fig.\ref{fig5}
but in this figure top-antitop spin is along the incoming beam
directions.

In our calculations phase space integrations have been performed by
GRACE \cite{grace} which uses a Monte Carlo routine.

\section{Conclusion}

In this paper we have studied the top spin correlations with the
unparticle effects in $e^{+}e^{-}$ collision. We calculate spin
dependent cross sections to obtain top and antitop spin asymmetry.
It is shown that existence of scalar or vector unparticle leads to a
significant deviation of the spin asymmetry from the its SM value.
Variations of the spin asymmetry are reflected by the angular
distribution of the top and antitop decay products. Therefore spin
correlations provide us useful information to test the new physics
beyond the SM.


\pagebreak

\begin{table}
\caption{Spin asymmetries of the top quark pair for the process
$e^{+}e^{-}\rightarrow t\overline{t}$ in the helicity basis.
$\Lambda_{U}=1 TeV$, $d_{U}=1.1$ and $\sqrt{s}=0.5$ TeV.
\label{tab1}}
\begin{ruledtabular}
\begin{tabular}{ccc}
SM & $\textit{A}_{SM}=-0.651$ \\
Scalar Unparticle & $\textit{A}_{S}=0.986$ \\
Vector Unparticle & $\textit{A}_{V}=-0.727$ \\
\end{tabular}
\end{ruledtabular}
\end{table}

\begin{table}
\caption{Spin asymmetries of the top quark pair for the process
$e^{+}e^{-}\rightarrow t\overline{t}$. Top quark spin axis is along
the positron beam direction and antitop quark spin axis is along the
electron beam direction. $\Lambda_{U}=1 TeV$, $d_{U}=1.1$ and
$\sqrt{s}=0.5$ TeV. \label{tab2}}
\begin{ruledtabular}
\begin{tabular}{ccc}
SM & $\textit{A}_{SM}=-0.939$ \\
Scalar Unparticle & $\textit{A}_{S}=0.657$ \\
Vector Unparticle & $\textit{A}_{V}=-1$ \\
\end{tabular}
\end{ruledtabular}
\end{table}

\end{document}